\pdfoutput=1
\documentclass[a4paper,aip,american,citeautoscript,floatfix,jcp,pdftex,superscriptaddress,%
longbibliography,%
reprint,twocolumn%
]{revtex4-1}

\usepackage{amsmath,amssymb}
\usepackage{color}
\usepackage[T1]{fontenc}
\usepackage{graphicx}
\usepackage[utf8]{inputenc}
\usepackage{microtype}
\usepackage{textcomp}
\usepackage[textsize=footnotesize]{todonotes}
\usepackage{xspace}
\usepackage{hyperref}
\usepackage[ams,todos]{CMImacros}
\usepackage{booktabs}

\newlength{\figwidth}
\newcommand{\cfeldesy}{\affiliation{Center for Free-Electron Laser Science, Deutsches Elektronen-Synchrotron DESY, Notkestrasse 85, 22607 Hamburg, Germany}}%
\newcommand{\uhhcui}{\affiliation{The Hamburg Center for Ultrafast Imaging, Universität Hamburg, Luruper Chaussee 149, 22761 Hamburg, Germany}}%
\newcommand{\cmiweb}{\homepage{https://www.controlled-molecule-imaging.org}}%
\newcommand{\ayemail}{\email{andrey.yachmenev@cfel.de}}%

\begin{document}

\title{RichMol: A general variational approach for rovibrational molecular dynamics in external electric fields}%

\author{Alec Owens}\cfeldesy\uhhcui

\author{Andrey Yachmenev}\ayemail\cmiweb\cfeldesy\uhhcui%

\date{\today}%

\begin{abstract}
A general variational approach for computing the rovibrational dynamics of polyatomic molecules in the presence of external electric fields is presented. Highly accurate, full-dimensional variational calculations provide a basis of field-free rovibrational states for evaluating the rovibrational matrix elements of high-rank Cartesian tensor operators, and for solving the time-dependent Schr\"odinger equation. The effect of the external electric field is treated as a multipole moment expansion truncated at the second hyperpolarizability interaction term. Our fully numerical and computationally efficient method has been implemented in a new program, RichMol, which can simulate the effects of multiple external fields of arbitrary strength, polarization, pulse shape and duration. Illustrative calculations of two-color orientation and rotational excitation with an optical centrifuge of NH$_3$ are discussed.
\end{abstract}

\maketitle

\section{Introduction}

 The ability to monitor and control molecular systems with strong tailored light sources has seen remarkable progress in recent years, particularly on the experimental side~\cite{Lemeshko:MP111:1648}. Generally speaking, theoretical predictions of molecule-field interactions tend to evade rigorous description and molecules are often treated as rigid rotors with field effects approximated by dipole and polarizability interactions, see e.g. Ref.~\onlinecite{Chang:CPC185:339}. Such approaches are valid for moderate field strengths and are capable of describing the rotational dynamics of quasi-rigid molecules in well-isolated vibrational states. For non-rigid polyatomic molecules this no longer applies, especially in the presence of strong electric fields which can couple several vibrational and rotational states, significantly changing the rovibrational molecular dynamics. It is desirable, therefore, to have more sophisticated theoretical approaches that simultaneously consider many coupled internal vibrational and rotational motions, and which incorporate nonlinear effects.

 In the last decade or so, a number of general-purpose variational codes for computing the rovibrational energy levels and transition probabilities of polyatomic molecules have been developed~\cite{Avila:JCP139:134114,Wang:JCP130:094101,
Matyus:JCP127:084102,Matyus:JCP130:134112,Yurchenko:JMS245:126,
Yachmenev:JCP143:014105,
Yurchenko:JCTC13:4368}. One of these is the nuclear motion program TROVE,~\cite{Yurchenko:JMS245:126,
Yachmenev:JCP143:014105,
Yurchenko:JCTC13:4368} which utilizes a completely numerical procedure to generate the Hamiltonian matrix and considers all interactions between rotational and vibrational motion. Along with its algorithmic efficiency, TROVE benefits from the use of molecular symmetry including non-Abelian symmetry groups,~\cite{Yurchenko:JCTC13:4368} curvilinear internal coordinates and the Eckart coordinate frame.~\cite{Yachmenev:JCP143:014105} The program continues to be used extensively for molecular line list production~\cite{Tennyson:IJQC117:92} and represents the current state-of-the-art in time-independent rovibrational calculations. With such a powerful computational tool, a natural extension of TROVE, or any other variational code for that matter, is towards the simulation of the rovibrational dynamics of molecules in the presence of external electric fields.

In this work, we present a general and computationally efficient approach for computing the effects of external electric fields, which are treated by multipole moment expansion of order up to the second hyperpolarizability interaction tensor. A fully numerical method is described for evaluating the rovibrational matrix elements of Cartesian tensor operators, and for solving the time-dependent Schr\"odinger equation. Our approach has been implemented in a new computer program RichMol, which utilizes the field-free rovibrational energies and wavefunctions from TROVE. To our knowledge, this is the first attempt to create a general and robust variational approach for computing electric field effects in polyatomic molecules with high accuracy. Illustrative calculations are presented on the ammonia molecule of two-color field-free orientation dynamics and rotational excitation with an optical centrifuge.

\section{Methodology}

 In the Born-Oppenheimer approximation, the effects of an external electric field can be described by a time-dependent potential $V(t)$ expanded in terms of molecular electric multipole moments as
\begin{eqnarray}\label{eq:vpot}
  V(t) &=& -\mu_A E_A(t) - \frac{1}{2}\alpha_{AB}E_A(t)E_B(t) \\ \nonumber
  &&- \frac{1}{6}\beta_{ABC}E_A(t)E_B(t)E_C(t) \\ \nonumber
  &&-\frac{1}{24}\gamma_{ABCD}E_A(t)E_B(t)E_C(t)E_D(t).
\end{eqnarray}
Here, $A,B,C,D$ are Cartesian indices denoting the $x,y,z$ axes in the laboratory frame and the summation over all Cartesian indices is implicitly assumed, $E_A(t)$ is the $A$ Cartesian component of the electric field vector, and $\mu_A$, $\alpha_{AB}$, $\beta_{ABC}$, and $\gamma_{ABCD}$ are the electronic contributions to the electric dipole, polarizability, first and second hyperpolarizability Cartesian multipole tensors, respectively. These tensors can be determined in the molecular frame from electronic structure calculations as functions of internal vibrational coordinates of the molecule. In general, they depend on the frequency of the applied electric field, however, for non-resonant field frequencies this dependence can usually be neglected.

 The rovibrational dynamics of a molecule in the presence of an electric field is described by a Hamiltonian composed of the field-free rovibrational Hamiltonian $H_{\rm rv}$ and the field interaction potential $V(t)$, i.e.
\begin{eqnarray}\label{eq:htot}
  H(t) = H_{\rm rv} + V(t).
\end{eqnarray}
The eigenvalues and eigenvectors of $H_{\rm rv}$, which are the rovibrational energies $E_{Jml}^{(\rm rv)}$ and wavefunctions $\Psi_{Jml}^{(\rm rv)}$, respectively, can be computed to a high degree of accuracy using any available variational code, such as TROVE. Here, $J$ is the quantum number of the total angular momentum operator, $m=-J,\ldots, J$ is the projection of $J$ onto the $z$-axis in the laboratory coordinate frame, and $l$ is a running number which identifies the rovibrational state. Note that in field-free calculations $m$ can be omitted as the $2J+1$ energy levels are degenerate.

 The wavefunctions $\Psi_{Jml}^{(\rm rv)}$ are usually represented as linear combinations of products of vibrational wavefunctions  $|v\rangle$ (solutions of the pure vibrational $J=0$ problem) and Wang-type symmetric top functions $|J,k,\tau,m\rangle$,
\begin{eqnarray}\label{eq:psi}
  \Psi_{Jml}^{(\rm rv)} = \sum_{v,k,\tau} c_{vk\tau}^{(Jl)} |v\rangle |J,k,\tau,m\rangle,
\end{eqnarray}
where $v$ denotes the vibrational state number, $k=0,\ldots, J$ is the projection of $J$ onto the $z$-axis in the molecular frame, and $\tau=0$ or 1 defines the rotational parity as $(-1)^\tau$. The eigenvalue coefficients $c_{vk\tau}^{(Jl)}$, identifiable by $J$ and $l$, are obtained for a set of rovibrational states by diagonalising a matrix representation of the rovibrational Hamiltonian, done separately for each value of $J$. For details of the field-free variational solution in TROVE the reader is referred to Refs.~\onlinecite{Yurchenko:JMS245:126,Yachmenev:JCP143:014105,
Yurchenko:JCTC13:4368,Tennyson:IJQC117:92,Yachmenev:JCP147:141101}.

 The field-free wavefunctions $\Psi_{Jml}^{(\rm rv)}$ can be used as a basis to represent the time-dependent solutions $\Phi(t)$ of the Schr\"odinger equation for the total Hamiltonian $H(t)$, i.e.
\begin{eqnarray}\label{eq:phi}
  \Phi(t) = \sum_{J,m,l} C_{Jml}(t) \Psi_{Jml}^{(\rm rv)},
\end{eqnarray}
where the time-dependent expansion coefficients $C_{Jml}(t)$ are obtained from solution of the time-dependent Schr\"odinger equation. Several numerical techniques have been developed for solving the time-dependent problem using the time discretization method~\cite{Leforestier:JCOP94:59,Feit:JCOP47:412,Bandrauk:jcp99:1185}. These techniques require multiple evaluations of matrix-vector products between the matrix representation of the time-dependent Hamiltonian, $\langle \Psi_{J'm'l'}^{(\rm rv)}|H(t)| \Psi_{Jml}^{(\rm rv)}\rangle$
= $E_{Jml}^{(\rm rv)}\delta_{J'J}\delta_{m'm}\delta_{l'l}$ + $\langle \Psi_{J'm'l'}^{(\rm rv)}|V(t)| \Psi_{Jml}^{(\rm rv)}\rangle$, and the vector of the expansion coefficients $C_{Jml}(t)$. To proceed, it is essential to develop a general and computationally efficient approach for calculating the $\Psi_{Jml}^{(\rm rv)}$-basis matrix elements of the Cartesian tensor operators in the expression for $V(t)$ of Eq.~\eqref{eq:vpot}. To facilitate the linear algebra operations, a factorization of matrix elements in terms of tensors of smaller dimension is highly beneficial.

 Derivation of the analytical expressions for the matrix elements of the laboratory frame Cartesian tensor operators in Eq.~(\ref{eq:vpot}) is a formidable task, particularly for tensors of high rank. Instead, we develop a simple numerical scheme for computing the matrix elements which provides the same efficiency as methods based on prederived analytical expressions of tensors of low rank. We will consider only electric field tensor operators, which all possess full permutational index symmetry. Hence, only symmetry-unique elements of the tensors will be kept in the form of a vector with indices denoted by $A$ and $\alpha$ for tensors in the laboratory and molecular frame, respectively.

 We define a generalized Cartesian tensor operator in the laboratory and molecular frame, $T_A^{(\rm LF)}$ and $T_\alpha^{(\rm MF)}$, of rank $\Omega$ with elements $A$ or $\alpha=x,y,z$ for $\Omega=1$, $A$ or $\alpha=xx,xy,xz,yy,yz,zz$ for $\Omega=2$, and so on. The relationship between $T_A^{(\rm LF)}$ and $T_\alpha^{(\rm MF)}$ is most easily established by transforming both into the irreducible spherical tensor form~\cite{Zare:AngularMomentum}, e.g., for the laboratory frame tensor this reads
\begin{eqnarray}\label{eq:tens1}
T_{\omega\sigma}^{(\rm LF)} = \sum_{A} U_{\omega\sigma,A}^{(\Omega)} T_{A}^{(\rm LF)},
\end{eqnarray}
where $T_{\omega\sigma}^{(\rm LF)}$ is the $\sigma$ component of the irreducible spherical tensor operator ($\sigma=-\omega,\ldots,\omega$) of rank $\omega=0,\ldots,\Omega$. The elements of matrix $U_{\omega\sigma,A}^{(\Omega)}$ ($\equiv U_{\omega\sigma,\alpha}^{(\Omega)}$) are well known for tensors of low rank, see e.g. Table~\ref{tab:umat} or Ref.~\onlinecite{Zare:AngularMomentum}. For spherical tensor operators of high rank, $U_{\omega\sigma,A}^{(\Omega)}$ can be generated from the spherical tensors of lower rank by a successive application of the angular momentum coupling rule, i.e.
\begin{eqnarray}\label{eq:umat}
U_{\omega\sigma,A}^{(\Omega)}
&=& \sum_{\sigma_1=-\omega_1}^{\omega_1}\sum_{\sigma_2=-\omega_2}^{\omega_2}
\langle\omega_1\sigma_1\omega_2\sigma_2|\omega\sigma\rangle \\ \nonumber
&\times&  U_{\omega_1\sigma_1,B}^{(\Omega_1)} U_{\omega_2\sigma_2,C}^{(\Omega_2)}.
\end{eqnarray}
Here, $\langle\omega_1\sigma_1\omega_2\sigma_2|\omega\sigma\rangle$ is the Clebsch-Gordan coefficient, $\Omega = \Omega_1 + \Omega_2$ and the Cartesian label $A=B\otimes C$, e.g., $A=yzx$ for $B=yz$ and $C=x$. As an example, in Table~\ref{tab:umat} elements of the matrix $U^{(2)}$ for the electric polarizability tensor are obtained from $U^{(1)}$ for the dipole moment using Eq.~(\ref{eq:umat}).

 Spherical tensor operators in the molecule-fixed $T_{\alpha}^{(\rm MF)}$ and laboratory-fixed $T_{A}^{(\rm LF)}$ frames are related by a Wigner rotation $D$-matrix~\cite{Zare:AngularMomentum}. From Eq.~(\ref{eq:tens1}), this relationship can be established for the Cartesian tensor,
\begin{eqnarray}\label{eq:tens2}
T_{A}^{(\rm LF)} =
\sum_{\omega=0}^{\Omega}\sum_{\sigma,\nu=-\omega}^{\omega}\sum_{\alpha}
[U^{(\Omega)}]^{-1}_{A,\omega\sigma} [D_{\sigma\nu}^{(\omega)}]^*
U^{(\Omega)}_{\omega\nu,\alpha} T_{\alpha}^{(\rm MF)},
\end{eqnarray}
where $[U^{(\Omega)}]^{-1}$ is the pseudo-inverse matrix of $U^{(\Omega)}$, the Wigner $D$-matrix depends on the rotational coordinates (Euler angles) and $T_{\alpha}^{(\rm MF)}$ is a function of the internal vibrational coordinates only. Thus, integration of $T_{A}^{(\rm LF)}$ in the product basis of rotational and vibrational functions, given by Eq.~\eqref{eq:psi}, can be carried out separately.

 This leads to the following  expression for the rovibrational matrix elements
\begin{eqnarray}\label{eq:tens3}
\langle \Psi_{J'm'l'}^{(\rm rv)} |T_{A}^{(\rm LF)}| \Psi_{Jml}^{(\rm rv)} \rangle &=&
\sum_{\omega=0}^{\Omega} \mathcal{M}_{A\omega}^{(J'm',Jm)}
\mathcal{K}_{\omega}^{(J'l',Jl)},
\end{eqnarray}
with
\begin{eqnarray}\label{eq:mmat}
\mathcal{M}_{A\omega}^{(J'm',Jm)} &=& (-1)^{m'}\sqrt{(2J'+1)(2J+1)} \\ \nonumber
&\times&\sum_{\sigma=-\omega}^{\omega} [U^{(\Omega)}]^{-1}_{A,\omega\sigma}
\left(\begin{array}{ccc}J&\omega&J'\\m&\sigma&-m'\end{array}\right),
\end{eqnarray}
and
\begin{eqnarray}\label{eq:kmat}
\mathcal{K}_{\omega}^{(J'l',Jl)} = \sum_{\substack{k',\tau',v'\\k,\tau, v}}
\left[c_{v'k'\tau'}^{(J'l')}\right]^* c_{vk\tau}^{(Jl)}
\sum_{\pm k',\pm k} \left[d_{k'}^{(\tau')}\right]^*d_{k}^{(\tau)} \\\nonumber
\times (-1)^{k'} \sum_{\sigma=-\omega}^{\omega}\sum_{\alpha}
\left(\begin{array}{ccc}J&\omega&J'\\k&\sigma&-k'\end{array}\right)
U^{(\Omega)}_{\omega\sigma,\alpha}\langle v' |T_{\alpha}^{(\rm MF)}| v\rangle,
\end{eqnarray}
where the Wang coefficients $d_{\pm k}^{(\tau)}$ define the symmetric-top functions $|J,k,\tau,m\rangle=d_{+k}^{(\tau)}|J,k,m\rangle+d_{-k}^{(\tau)}|J,-k,m\rangle$. The matrix elements $\mathcal{K}_{\omega}$ are independent of the laboratory frame and can be precomputed and stored for rovibrational states of interest. Importantly, by prescreening the non-zero elements in the sum of products of the $3j$-symbols with the elements of the transformation matrix $U^{(\Omega)}$, the rigorous selection rules are automatically fulfilled. The number of arithmetic operations required for evaluating the computationally expensive double sum over the $\{k'\tau'v'\}$ and $\{k\tau v\}$ quanta in Eq.~\eqref{eq:kmat} is reduced to the same number as if the expressions were prederived analytically. Depending on the rank of the tensor $\Omega$ and irreducible representation $\omega$, the elements of $\mathcal{K}_{\omega}$ and $\mathcal{M}_{\omega}$ are either purely real or imaginary numbers. Complex-valued arithmetic is therefore unnecessary if the effect of the imaginary unit $i$ in real-valued operations is properly taken care of.

The vibrational matrix elements $\langle v' |T_{\alpha}^{(\rm MF)}| v\rangle$ in Eq.~\eqref{eq:kmat}
are computed in TROVE by expanding the electric field tensors as power series
in terms of the coordinates describing molecular vibrations.
TROVE implements expansions around one equilibrium geometry for quasi-rigid vibrations,
as well as expansions around multiple points on a grid of geometries describing non-rigid vibration~\cite{Yurchenko:JMS245:126,Yachmenev:JCP143:014105}.
Some variational approaches use a pointwise representation of operators
on multi-dimensional grids in terms of the vibrational coordinates~\cite{Avila:JCP139:134114,Wang:JCP130:094101,
Matyus:JCP127:084102,Matyus:JCP130:134112},
in this case the vibrational matrix elements are computed using numerical integration
by quadratures.

 For time propagation of the wavepacket $\Phi(t)$, the bottleneck operation is evaluating the matrix-vector products between the matrix representation of the total Hamiltonian $H(t)$ and the wavepacket coefficients $C_{Jml}(t)$. The matrix representation of $H_{\rm rv}$ is diagonal with the elements $E_{Jml}^{(\rm rv)}\delta_{J'J}\delta_{m'm}\delta_{l'l}$, whilst the matrix elements of the field potential $V(t)$ at a time $t$ can be written as
\begin{eqnarray}
\langle \Psi_{J'm'l'}^{(\rm rv)} |&V(t)&| \Psi_{Jml}^{(\rm rv)} \rangle = \\ \nonumber
&=& \sum_{n}f_n\sum_{\omega=0}^{\Omega_n}\tilde{\mathcal{M}}_{\omega,n}^{(J'm',Jm)}(t)
\mathcal{K}_{\omega,n}^{(J'l',Jl)}
\end{eqnarray}
where
\begin{eqnarray}
\tilde{\mathcal{M}}_{\omega,n}^{(J'm',Jm)}(t)
= \sum_{A} \mathcal{M}_{A\omega,n}^{(J'm',Jm)}E_A(t).
\end{eqnarray}
We have introduced the index $n=1,2,\ldots$ to distinguish between different electric multipole tensor operators in the expansion of $V(t)$ in Eq.~\eqref{eq:vpot}, with $f_n$ being the respective constant prefactors (e.g. $f=-1/2$ for the polarizability). Since different tensor operators have different rovibrational selection rules, it is more efficient to compute the matrix-vector products separately for each operator in the multipole expansion and sum the results at the last operation.

 Evaluation of the matrix-vector product of the matrix representation of $V(t)$ and the eigenvector coefficients $C_{Jml}(t)$, that is
\begin{eqnarray}
h_{J'm'l'} = \sum_{J,m,l} \langle \Psi_{J'm'l'}^{(\rm rv)} |V(t)| \Psi_{Jml}^{(\rm rv)} \rangle C_{Jml}(t)
\end{eqnarray}
is best carried out in two steps
\begin{eqnarray}
&&F_{\omega,n}^{(J'l',Jm)} = \sum_{l} \mathcal{K}_{\omega,n}^{(J'l',Jl)} C_{Jml}(t) \\
h_{J'm'l'} &=& \sum_{n}f_n\sum_{\omega}\sum_{J,m}F_{\omega,n}^{(J'l',Jm)}
\tilde{\mathcal{M}}_{\omega,n}^{(J'm',Jm)}(t).
\end{eqnarray}
This reduces the operation count by a factor of $(2J_{\rm max}+1)$ for the first summation, and by a factor of $l_{\rm max}$ for the second summation. Here, $J_{\rm max}$ is the maximal value of the quantum number $J$, and $l_{\rm max}$ is the maximal number of rovibrational functions (with the same $J$) spanned by the basis set.

The presented method, including solution of the time-dependent Schr\"odinger equation, has been implemented in a new computer program RichMol, which has been interfaced with TROVE. In the next section we present illustrative applications involving the ammonia molecule in different electric field scenarios.

\begin{table}[htb]
\caption{Nonzero elements of matrix $U^{(\Omega)}_{\omega\sigma,\alpha}$
in Eq.~(\ref{eq:umat}).
Summation over all index-symmetric elements is performed, e.g.,
$U^{(\Omega)}_{\omega\sigma,xy}=U^{(\Omega)}_{\omega\sigma,xy}+U^{(\Omega)}_{\omega\sigma,yx}$ .\label{tab:umat}}
\tabcolsep=4pt
\renewcommand{\arraystretch}{1.5}
\begin{tabular}{cccccccc}
\cmidrule(l){2-7}
\multicolumn{8}{c}{Dipole moment $\Omega=1$} \\
&&$(\omega,\sigma)$ && \multicolumn{3}{c}{$\alpha$} \\
\cmidrule(l){4-7}
&& && $x$ & $y$ & $z$ \\
&&$(1,-1)$ && $\frac{1}{\sqrt{2}}$ & $-\frac{i}{\sqrt{2}}$ & 0 \\
&&$(1,0)$ && 0 & 0 & 1\\
&&$(1,1)$ && $-\frac{1}{\sqrt{2}}$ & $-\frac{i}{\sqrt{2}}$ & 0 \\
\cmidrule(l){1-8}
\multicolumn{8}{c}{Polarizability $\Omega=2$} \\
$(\omega,\sigma)$ && \multicolumn{6}{c}{$\alpha$} \\
\cmidrule(l){2-8}
 && $xx$ & $xy$ & $xz$ & $yy$ & $yz$ & $zz$ \\
$(0,0)$ && $-\frac{1}{\sqrt{3}}$ & 0 & 0 & $-\frac{1}{\sqrt{3}}$ & 0 & $-\frac{1}{\sqrt{3}}$ \\
$(2,-2)$ && $\frac{1}{2}$ & $-i$ & 0 & $-\frac{1}{2}$ & 0 & 0 \\
$(2,-1)$ && 0 & 0 & 1 & 0 & $-i$ & 0 \\
$(2,0)$ && $-\frac{1}{\sqrt{6}}$ & 0 & 0 & $-\frac{1}{\sqrt{6}}$ & 0 & $\sqrt{\frac{2}{3}}$ \\
$(2,1)$ && 0 & 0 & -1 & 0 & $-i$ & 0 \\
$(2,2)$ && $\frac{1}{2}$ & $i$ & 0 & $-\frac{1}{2}$ & 0 & 0 \\
\cmidrule(l){1-8}
\end{tabular}
\end{table}

\section{Applications}

\subsection{Analytical representation of electric field tensors}

 To compute the vibrational matrix elements $\langle v'|T_{\alpha}^{(\rm MF)}|v\rangle$ of the molecular frame electric field tensors in Eq.~\eqref{eq:kmat}, the property surfaces generated from the electronic structure calculations need to be represented by analytical functions.
 For methods that use numerical integration, analytical representations of the electric field tensors are not essential but can be hugely rewarding in terms of reducing the number of costly electronic structure calculations.
 If possible, it is also desirable to take into account the symmetry of the molecule by using symmetry-adapted combinations of tensor elements.

 To simplify the symmetrization procedure, it is convenient to first transform the Cartesian tensors into a coordinate system with more straightforward symmetry properties. One of the methods is to employ a coordinate system spanned by unit vectors along the molecular bonds, i.e. the molecular bond (MB) frame~\cite{Yurchenko:JPCA113:11845}. In the MB frame, permutations of identical atoms lead to permutations of the molecular bonds, and hence permutations of the MB unit vectors and corresponding tensor projections. For the ammonia molecule, the MB unit vectors become linearly dependent at planar configurations and one extra vector pointing along the trisector (three-fold symmetry axis) has to be introduced to make the transformation invertible.

 To illustrate the idea, we consider NH$_3$ in the {\bf D}$_{3h}$(M) molecular symmetry group and aim to construct a symmetry-adapted representation of the fully index-symmetric hyperpolarizability tensor $\boldsymbol\beta$. The MB frame is spanned by three unit vectors, one along each of the N--H$_i$ ($i=1,2,3$) bonds, and one unit vector pointing along the trisector. This coordinate system has been employed previously for NH$_3$ to represent the electric dipole moment (first-rank tensor)~\cite{Yurchenko:JPCA113:11845} and the $N$-atom electric field gradient (second-rank tensor)~\cite{Yachmenev:JCP147:141101}.

 The transformation matrix ${\bf S}(4\times 3)$ from the Cartesian to the MB frame is defined in terms of the Cartesian coordinates of the nuclei ${\bf r}_{{\rm H}_i}$ and ${\bf r}_{{\rm N}}$ such that
\begin{eqnarray}\label{eq:tmat}
S_{i\alpha} &=& \left( {\bf r}_{{\rm H}_i}-{\bf r}_{{\rm N}}\right) / r_{{\rm NH}_i},~~~\text{for}~i=1,2,3, \\ \label{eq:tmat2}
S_{4\alpha} &=& ({\bf S}_1\times {\bf S}_2 + {\bf S}_2\times {\bf S}_3 + {\bf S}_3\times {\bf S}_1 ),
\end{eqnarray}
where $r_{{\rm NH}_i}$ is the N--H$_i$ bond distance, and the ${\bf S}_4$ vector is assumed to be normalized. The transformation of $\boldsymbol\beta$ reads
\begin{eqnarray}
\beta_{ijk} &=& \sum_{\alpha,\beta,\gamma=x,y,z} S_{i\alpha} S_{j\beta} S_{k\gamma}
\beta_{\alpha\beta\gamma},
\end{eqnarray}
where $\alpha,\beta,\gamma=x,y$, or $z$ of the Cartesian frame and $i,j,k=1,2,3$, or 4 denote the unit vectors of the MB frame. The symmetry properties of the projections $\beta_{ijk}$ are alike for the N--H bonds and the trisector. For example, the $C_3$ rotation symmetry operation, which is isomorphic to the nuclear permutation (H$_1$,H$_2$,H$_3$), transforms the element $\beta_{124}$ into $\beta_{234}$. The operation of improper rotation $S_3$, which is isomorphic to $C_3$ followed by inverting the direction of the trisector, transforms $\beta_{124}$
to $-\beta_{234}$. By using the projection operator method, as described for example in Ref.~\onlinecite{Bunker:MolecularSymmetry}, all possible symmetry-adapted combinations of ${\beta}_{ijk}$ can be reconstructed.

 It follows, from classification with respect to the irreducible representations $D^{j\pm}$ ($j$ is weight and $\pm$ is parity) of the full rotational group, that $\boldsymbol\beta$ transforms as $D^{(1-)}\oplus D^{(3-)}$~\cite{Andrews:SpecActA46:871}. And by mapping $D^{j\pm}$ onto the corresponding representations of {\bf D}$_{3h}$(M), $D^{(1-)}$ and $D^{(3-)}$ transform as $A_2''+E'$ and $A_1'+A_2'+A_2''+E'+E''$, respectively. The symmetry-adapted combinations thus read
\begin{eqnarray}\label{eq:beta_sym}
\beta_1^{(A_1')} &=& \frac{1}{\sqrt{3}}(\beta_{111} + \beta_{222} + \beta_{333}) \\ \nonumber
\beta_2^{(A_2')} &=& \frac{1}{\sqrt{6}}(\beta_{112} - \beta_{113} - \beta_{122} + \beta_{133} + \beta_{223} - \beta_{233}) \\ \nonumber
\beta_3^{(A_2'')} &=& \frac{1}{\sqrt{3}}(\beta_{114} + \beta_{224} + \beta_{334}) \\ \nonumber
\beta_4^{(A_2'')} &=& \beta_{444} \\ \nonumber
\beta_5^{(E_a')} &=& \frac{1}{2\sqrt{3}}(-\beta_{112} + 2\beta_{113} - \beta_{122} - \beta_{133} + 2\beta_{223} - \beta_{233}) \\ \nonumber
\beta_5^{(E_b')} &=& \frac{1}{2}(\beta_{112} - \beta_{122} - \beta_{133} + \beta_{233}) \\ \nonumber
\beta_6^{(E_a')} &=& \frac{1}{\sqrt{6}}(\beta_{144} + \beta_{244} - 2\beta_{344}) \\ \nonumber
\beta_6^{(E_b')} &=& \frac{1}{\sqrt{2}}(\beta_{144} - \beta_{244}) \\ \nonumber
\beta_7^{(E_a'')} &=& \frac{1}{\sqrt{6}}(2\beta_{124} - \beta_{134} - \beta_{234}) \\ \nonumber
\beta_{7}^{(E_b'')} &=& \frac{1}{\sqrt{2}}(\beta_{134} - \beta_{234}).
\end{eqnarray}
Here, the $E_a$ and $E_b$ symmetry components of the doubly degenerate representations $E'$ and $E''$ are connected by a simple orthogonal transformation and can be parametrized by one set of constants. In total, seven symmetry-unique combinations are sufficient to describe the hyperpolarizability tensor for NH$_3$, which are individually represented by analytical functions determined from fitting to the electronic structure data points. In variational calculations, the symmetry-adapted combinations are transformed back to the Cartesian system (e.g. Eckart system) by applying the inverse transformation given by the ${\bf S}^{-1}$ matrix.

 For simulations of NH$_3$, the electric dipole moment, polarizability, first and second hyperpolarizability tensors were computed {\it ab initio} at the CCSD/aug-cc-pVTZ~\cite{Dunning:JCP90:1007,Kendall:JCP96:6796} level of theory in the frozen-core approximation. Calculations were performed using the response-theory coupled-cluster approach~\cite{Halkier:JCP107:849,Christiansen:JCP108:2801,Haettig:CPL269:428,Haettig:CPL282:139},
as implemented in the Dalton program package~\cite{Aidas:WIRE4:269}. The symmetry-unique tensor representations, obtained via the procedure outlined above, were parametrized using sixth order symmetry-adapted power series expansions through least-squares fittings. The same coordinates have been used in a previous study to represent the dipole moment of NH$_3$~\cite{Yurchenko:JPCA113:11845}. The values of the optimized parameters and the Fortran~90 functions together with the reference data points used for the fitting are provided in the supplementary material. The field-free basis of rovibrational wavefunctions was generated in TROVE using the potential energy surface and computational setup of Ref.~\onlinecite{Yurchenko:MNRAS413:1828}.

\subsection{Two-color orientation}\label{sec:two-color}

Laser-assisted alignment and orientation of gas-phase molecules is particularly
important in laser induced diffraction experiments with electrons and x-rays
\cite{Bisgaard:Science323:1464,Yang:NatComm7:11232,Kuepper:PRL112:083002}.
Over the years, several techniques have been developed for aligning and orienting
molecules in space. Orientation methods include the combination of electrostatic fields
and non-resonant laser excitation \cite{Holmegaard:PRL102:023001,Ghafur:NatPhys5:289},
linearly polarized laser pulses with $45^\circ$-skewed mutual polarization
\cite{Fleischer:NJP11:105039,Kitano:PRL103:223002,Yachmenev:PRL117:033001},
THz pulses \cite{Fleischer:PRL107:163603,Kitano:PRA84:053408,Egodapitiya:PRL112:103002},
and two-color laser fields \cite{De:PRL103:153002,Frumker:PRL109:113901}.
To demonstrate the capabilities of TROVE and RichMol in computing nonlinear field effects,
we perform simulations of the impulsive two-color orientation of NH$_3$,
where the interaction with the laser field occurs through the molecular polarizability
and hyperpolarizability tensors.

The two-color laser field is modeled by the function
\begin{eqnarray}
  E(t) = E_0e^{-4\ln2\frac{(t-t_0)^2}{\tau^2}}\left[\cos(\omega_1 t)+\cos(\omega_2 t)\right],
\end{eqnarray}
where $E_0$ is the field amplitude, the carrier frequencies are kept fixed at $\omega_1=400$~nm and $\omega_2=800$~nm,
and the pulse time profile is described by a Gaussian function with a maximum value at $t_0$ and a full width at half maximum (FWHM) of $\tau$.
The field is polarized along the laboratory-fixed $Z$ direction and the external field potential in Eq.~(\ref{eq:vpot}) includes polarizability
and hyperpolarizability interaction terms. For highly oscillatory fields,
the contribution from the dipole interaction averages out to zero.

 The time dependent wavepacket $\psi(t)$ is obtained in the basis of field-free rovibrational
states, where all stretching and bending quanta are fixed at zero,
the inversion motion quantum number $\nu_2^{\pm}$ ranges from $0,\ldots,2$, and all rotational quanta for $J\leq 20$ are included.
Calculations were performed on an initial ground state with mixed ($+$ and $-$) inversion parity, that is $\psi(0)=1/\sqrt{2}(|J,k,m,\nu_2^+\rangle+|J,k,m,\nu_2^-\rangle)$
with $J=1$, $k=1$, $m=0,\pm 1$ and $\nu_2=0$.
The orientation is characterized by the time-dependent expectation
value $\langle\psi(t)|\cos\theta|\psi(t)\rangle$, where $\theta$ is the Euler angle.

The time-dependent wavepacket coefficients are obtained by numerical
solution of the time-dependent Schr\"odinger equation using the split-operator method. The time
evolution of the wavepacket from time $t'\rightarrow t$ is described by the time-evolution
operator $U(t,t')$ such that $\psi(t)=U(t,t')\Psi(t')$, where $U(t,t')$ is evaluated as
\begin{eqnarray}
  U(t,t') &=& \exp\left[-i\frac{\Delta t}{2\hbar} H_{\rm rv}\right] \cdot \exp\left[-i\frac{\Delta t}{\hbar} V\left(\frac{t+t'}{2}\right)\right] \\ \nonumber
  &\cdot& \exp\left[-i\frac{\Delta t}{2\hbar} H_{\rm rv}\right],
\end{eqnarray}
with $\Delta{t}=t-t'$.
The exponential of the matrix representation of $V(t)$ is computed using the iterative
approximation based on Krylov subspace methods, as implemented in the Expokit computational package~\cite{Sidje:TOMS24:130}.
For the evaluated field configurations (see above) we found that a
discrete time step $\Delta{t}$ in the range 1--10~fs worked well.

\begin{figure}
\centering
\includegraphics[width=\columnwidth]{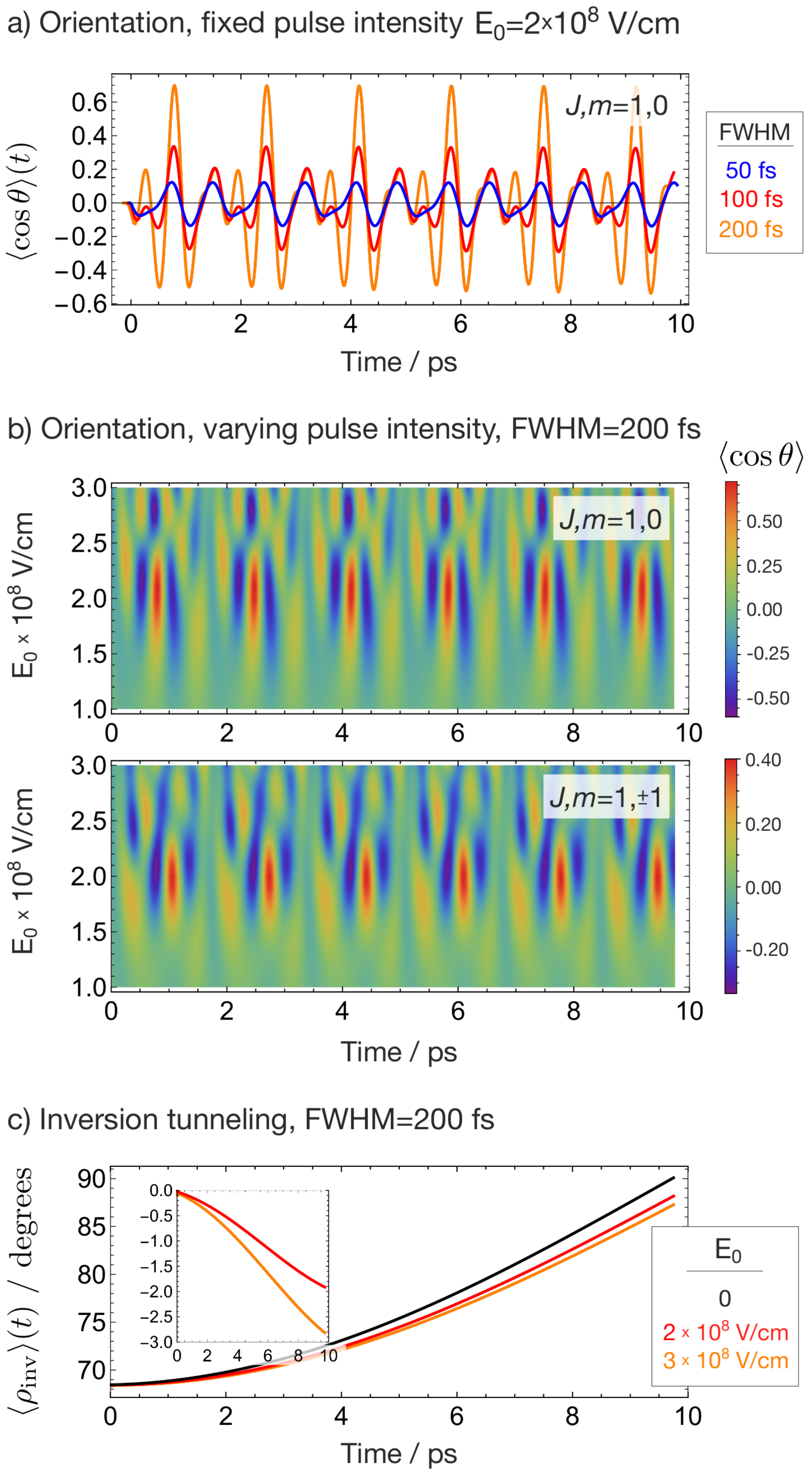}
\caption{\label{fig1}
Two-color impulsive orientation simulations for NH$_3$.
In panel (a), the orientation $\langle\psi(t)|\cos\theta|\psi(t)\rangle$ is shown for different pulse durations (FWHM).
In panel (b), the orientation is plotted for different pulse intensities ($E_0$),
and in panel (c), the time evolution of the expectation value of the inversion coordinate $\langle\psi(t)|\rho_{\rm inv}|\psi(t)\rangle$ is plotted for different pulse intensities.
The inset in panel (c) shows the inversion dynamics with subtracted natural (field-free) tunneling. }
\end{figure}

In Fig.~\ref{fig1}.a, revivals of the time-dependent orientation of NH$_3$ after the two-color
laser pulse are shown.
In each plot, the origin is set at the center of the Gaussian pulse $t_0$.
As expected, increasing the pulse duration (FWHM) significantly enhances the degree of alignment.
However, by increasing the pulse intensity and duration, the wavepacket is substantially depleted due to ionization, which is discussed in detail in Ref.~\onlinecite{Spanner:PRL109:113001}.
The periodic behavior of the orientation dynamics follows the quantum rotational revival pattern
with a rotational period $T_{\rm rot}=h/(2B_e)=1.67$~ps, where $B_e\sim$10.0~cm$^{-1}$ is the rotational constant of NH$_3$.

The influence of the laser field intensity on the orientation dynamics is
plotted in Fig.~\ref{fig1}.b for two initial states $\psi_{110}(0)$ and $\psi_{11\pm 1}(0)$.
For stronger field intensities, despite neglecting ionization depletion,
the orientation revivals become weaker due to the significant population
of high rotationally excited states.
The effect of the two-color pulse on the inversion tunneling dynamics of NH$_3$ is plotted in Fig.~\ref{fig1}.c.
As can be seen, the temporal evolution of the expectation value of the inversion coordinate $\rho_{\rm inv}$
($\rho_{\rm inv}=90^\circ$ at planar geometry)
does not notably change after the laser pulses and almost follows the field-free tunneling path.

\subsection{Rotational excitation with an optical centrifuge}

\begin{figure}
\centering
\includegraphics[width=\columnwidth]{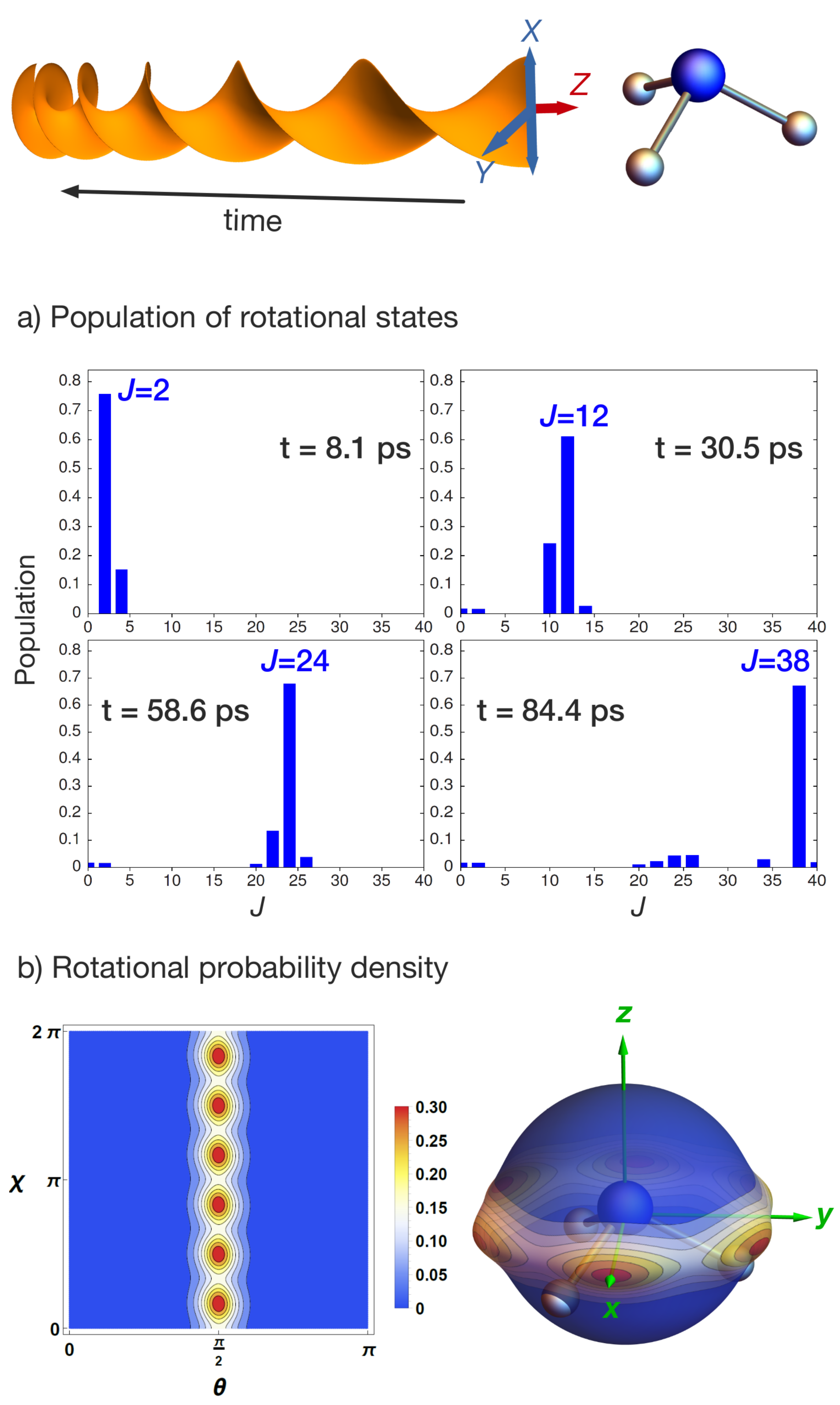}
\caption{\label{fig:oc_nh3}Optical centrifuge simulations of NH$_3$. In panel (a), the wavepacket population $\braket{\psi(0)}{\psi(t)}$ is shown as a function of $J$ at different times during the centrifuge pulse. Note that the rotational energy levels shown have the quantum number $m=-J$. In panel (b), the rotational probability density $P(\theta,\chi)$ is plotted as a function of the Euler angles $\theta$ and $\chi$. The two-dimensional plot has been projected on a three-dimensional sphere to illustrate how NH$_3$, and the molecule-fixed coordinate system, is oriented relative to the axes of rotation.}
\end{figure}

 The preparation of molecules in highly excited rotational states has seen a range of innovative methods developed in recent years~\cite{Ohshima:IRPC29:619,Milner:ACP159:395}. One of these is the optical centrifuge~\cite{Karczmarek:PRL82:3420,Villeneuve:PRL85:542}, which is a non-resonant, linearly polarized laser pulse that undergoes accelerated rotation along the direction of propagation (see Fig.~\ref{fig:oc_nh3}). The complexity of the experimental setup has limited the number of studies of centrifuged molecules, but molecular super-rotors, as they are known in the literature, are highly interesting objects for scattering~\cite{Mullin:JPCA119:12471}, spectroscopy~\cite{Yuan:FD150:101} and dynamics~\cite{Milner:PRX5:031041}. Given that only a handful of theoretical work on diatomic and linear triatomic molecules has been reported~\citep{Karczmarek:PRL82:3420,Spanner:JCP115:8403,Hasbani:JCP116:10636,Milner:JCP147:124202}, TROVE and RichMol provide a unique opportunity to explore new polyatomic molecules in an optical centrifuge. Furthermore, this can be done in a fully quantum mechanical manner.

 Simulations of NH$_3$ employed a field-free basis set containing rotational states in the ground vibrational state up to $J=40$, including both inversion-split state components. An optical centrifuge was applied for a duration of $t=84.4\,$ps along the laboratory-fixed $z$ axis and was represented by the following expression,
\begin{eqnarray}\label{eq:oc}
E(t) = E_0\cos(\omega t)\left[ {\bf e}_x\cos(\beta t^2) + {\bf e}_y\sin(\beta t^2) \right].
\end{eqnarray}
Here, $E_0=1.6\times 10^8\,$V/cm is the field amplitude,
$\beta=20\,$cm$^{-2}$ is the acceleration of circular rotation, and the carrier frequency of the field $\omega=800\,$nm.
Since the electric field is off-resonant, highly oscillating and not very strong,
the electric field potential in Eq.~(\ref{eq:vpot}) requires the polarizability
interaction term only.

 In Fig.~\ref{fig:oc_nh3}.a, snapshots of the wavepacket population $\braket{\psi(0)}{\psi(t)}$ during the centrifuge pulse have been plotted. Here, $\psi(t)$ is the wavepacket at time $t$ and $\psi(0)=|J,k,m,\nu_2^\pm\rangle=|0,0,0,0^-\rangle$ is the wavepacket at $t=0\,$ps. We see that NH$_3$ steadily climbs the rotational ladder through $\Delta J=2$, $\Delta m=-2$ rotational Raman transitions and by $t=84.4\,$ps, the dominant contribution to the wavepacket (67\%) is from the $J=38$, $m=-38$ state.  Also displayed in Fig.~\ref{fig:oc_nh3}.b is the rotational probability density function $P(\theta,\chi)=\int {\rm d}V {\rm d}\phi \,\psi(t)^{\ast}\psi(t) \sin \theta$, which provides information on the orientation of the molecule relative to the axes of rotation. The Euler angles are denoted by $\theta,\chi,\phi$ and ${\rm d}V$ is the volume element associated with the vibrational coordinates. By the end of the pulse, six ``islands'' have emerged which correspond to stable rotation axes,
 that are perpendicular to the $C_3$ molecular symmetry axis.
 The embedding of NH$_3$ into the Bloch sphere in Fig.~\ref{fig:oc_nh3}.b is used solely for illustrative purposes.
 In fact, NH$_3$ can exist simultaneously in both lower (shown on the figure) and upper
 pyramidal structures, since the optical centrifuge does not break the {\bf D}$_{\rm 3h}$(M)
 symmetry of the initial wavepacket (stationary state).

\begin{figure}
 \centering
 \includegraphics[width=\columnwidth]{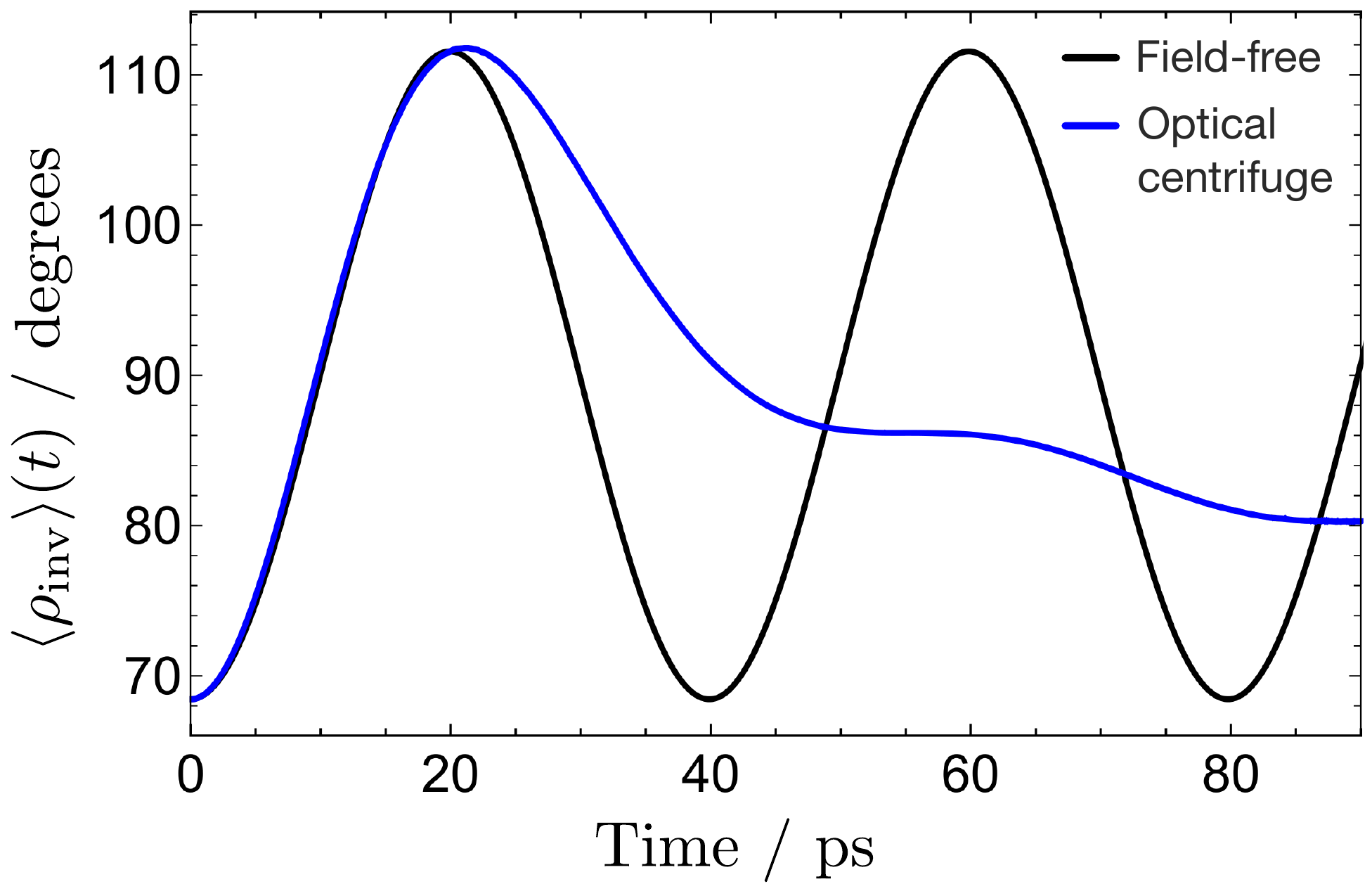}
 \caption{\label{fig3}
 The effect of rotational excitation with an optical centrifuge on the inversion tunneling dynamics in NH$_3$,
 computed as the time-dependent expectation value $\langle\psi(t)|\rho_{\rm inv}|\psi(t)\rangle$ of the inversion motion coordinate $\rho_{\rm inv}$.}
 \end{figure}

 To investigate the influence of the centrifuge rotation excitation
 on the inversion tunneling dynamics of NH$_3$, we have computed the temporal evolution
 of the expectation value of the inversion coordinate $\langle\psi(t)|\rho_{\rm inv}|\psi(t)\rangle$.
 We choose a mixed inversion parity initial wavepacket, i.e.
 $\psi(0)=1/\sqrt{2}(|J,k,m,\nu_2^+\rangle+|J,k,m,\nu_2^-\rangle)$, see Sec.~\ref{sec:two-color}.
 The results are displayed  in Fig.~\ref{fig3}. In the absence of an external field, the
 molecule tunnels between the two minima at $\rho_{\rm inv}\approx 68^\circ$ and $112^\circ$
 in about 20~ps.
 When the optical centrifuge is applied and the molecule starts to populate states with higher $J$,
 the inversion motion gradually slows down and becomes suppressed with a mean
 value at about $90^\circ$, which corresponds to a planar structure of NH$_3$.
 In classical terms, as the molecule rotates faster around one of the axes
 perpendicular to the $C_3$ symmetry axis of NH$_3$ (see Fig.~\ref{fig:oc_nh3}.b),
 centrifugal forces pull the hydrogen atoms towards a planar structure to minimize the energy.

\section{Conclusions}

 To our knowledge, this work represents the first attempt at a general variational approach for computing electric field effects in polyatomic molecules with high accuracy. Our method utilizes a basis set of field-free rovibrational states obtained from the nuclear motion code TROVE. The rovibrational matrix elements of the Cartesian tensor operators are evaluated in this basis, which is also employed for time-dependent simulations. External field effects are treated by multipole moment expansion with order up to the second hyperpolarizability interaction term. Our fully numerical and computationally efficient method has been implemented in a new program, RichMol. To illustrate the robustness of our approach, example calculations were presented on NH$_3$ of two-color orientation and rotational excitation with an optical centrifuge.

 TROVE and RichMol provide a computational tool capable of satisfying the high accuracy demands of modern experiment. With such an approach, a complete quantum mechanical description of the rovibrational dynamics in the presence of external electric fields is possible. Quantitative predictions, external field parameters, and time scales can all be obtained so that realistic experiments can be designed, but also interpreted. Although powerful there are limitations, namely that with current computational resources, variational calculations are only possible on molecules with less than 8--10 atoms. Extending to larger systems will require the use of reduced-dimensional models, but as long as these are carefully chosen there should be no loss of predictive power.

\section*{Supplementary material}

See the supplementary material for the expansion parameters and Fortran~90 functions to construct the electric dipole moment, polarizability, first and second hyperpolarizability tensors of NH$_3$.

\begin{acknowledgments}
The authors are grateful to Sergey Yurchenko for many valuable discussions. The computer program RichMol was started during the FP7-MC-IEF project 629237 ``Rotationally Induced Chirality in Molecules'' and we thank Ahmed Al-Refaie for suggesting the RichMol acronym.
Besides DESY, the authors also acknowledge support from the excellence cluster ``The Hamburg Center for Ultrafast Imaging---Structure, Dynamics and Control of Matter at the Atomic Scale'' of the Deutsche Forschungsgemeinschaft (CUI, DFGEXC1074), and the COST action MOLIM No. CM1405.
\end{acknowledgments}

\bibliography{string,cmi}
\end{document}